\documentclass[12pt]{iopart}
\usepackage{iopams}

\usepackage{graphicx}
\usepackage{dcolumn}
\usepackage{bm}
\def\cucox{Cu$_x$Co$_{1-x}$Fe$_2$O$_4$\ }
\def\cuco{Cu$_{0.5}$Co$_{0.5}$Fe$_2$O$_4$\ }

\def\SAB{{\it Sens.\ Actuators\ B:\ Chem\ }}
\def\JMSL{{\it J.\ Mater.\ Sc.\ Lett.\ }}
\def\MSEB{{\it Mater.\ Sc.\ Eng.\ B\ }}
\def\JMS{{\it J.\ Mater.\ Sc.\ }}
\def\JMSE{{\it J.\ Mater.\ Sc.:\ Mater.\ in Elec.\ }}
\def\JNPR{{\it J.\ Nanop.\ Research\ }}
\def\JMR{{\it J.\ Mater.\ Res.\ }}
\def\PhB{{\it Physica\ B\ }}
\def\MCP{{\it Mater.\ Chem.\ Phys.\ }}
\def\JMPT{{\it J.\ Mater.\ Processing Tech.\ }}
\def\MLett{{\it Mater.\ Lett.\ }}
\def\PAC{{\it Pure\ Appl.\ Chem.\ }}
\def\JNN{{\it J.\ Nano.\ Nanotech.\ }}

\begin{document}

\title[Magnetic and humidity sensing properties of \cucox]{Magnetic and humidity sensing properties of nanostructured \cucox synthesized by auto combustion technique}
\author{S. Muthurani$^1$,  M. Balaji$^1$, Sanjeev Gautam$^{2,3}$, Keun Hwa Chae$^2$, J.H. Song$^2$, D. Pathinettam Padiyan$^1$, K. Asokan$^4$}
\address{$^1$Department of Physics, Manonmaniam Sundaranar University, Tirunelveli, Tamilnadu, India}
\address{$^2$Nano Analysis Center, Korea Institute of Science and Technology, Seoul 136-791, Republic of Korea}
\address{$^3$Pohang Accelerator Laboratory, Pohang University of Science and Technology, Pohang -790 784, Republic of Korea}
\address{$^4$Inter-University Accelerator Centre, Aruna Asaf Ali Marg, New Delhi-110067 India.}
\ead{\mailto{sgautam71@kist.re.kr(S.Gautam)}}
\begin{abstract}
Magnetic nanomaterials (23-43 nm) of \cucox (x = 0.0, 0.5 and 1.0) were synthesized by auto combustion method. The crystallite sizes of these materials were calculated from X-ray diffraction peaks. The band observed in Fourier transform infrared spectrum near 575 cm$^{-1}$ in these samples confirm the presence of ferrite phase. Conductivity measurement shows the thermal hysteresis and demonstrates the knee points at 475$^o$C, 525$^o$C and 500$^o$C for copper ferrite, cobalt ferrite and copper-cobalt mixed ferrite respectively. The hystersis M-H loops for these materials were traced using the Vibrating Sample Magnetometer (VSM) and indicate a significant increase in the  saturation magnetization (M$_s$)  and remanence (M$_r$) due to the substitution of Cu$^{2+}$ ions in cobalt ferrite, while the intrinsic coercivity (H$_c$) was decreasing.  Among these ferrites, copper ferrite exhibits highest sensitivity for humidity.
\end{abstract}
\vspace{2pc}
\submitto{\NT}
\maketitle

\section{Introduction}\label{intro}
The increased concern about environmental protection led to the development in sensors field.  Apart from the technological importance ferrite materials have shown advantages in the field of sensors due to its mechanical strength, resistance to chemical attack and stability.  Ferrites have spinel structure, which is mainly used in gas \cite{r1,r2,r3}, stress \cite{r4} and humidity \cite{r5} sensors.  Humidity sensors are potentially in demand in industries like cloth driers, air coolers, broiler forming, cereal stocking and medical field \cite{r6}.  Humidity sensors based on the metal oxide materials have advantages such as low cost, simple construction and ease of placing the sensor in the operating environment.  The ability of a metal oxide to sense the presence of water molecules depends on the interaction between water molecules and surface of the metal oxide i.e. the reactivity of its surface.  The reactivity depends on the composition and morphological structure, which depends on the preparation procedure.  Ferrites can be prepared by sol-gel method \cite{r7}, co-precipitation method \cite{r8}, hydrothermal method \cite{r1}, milling \cite{r9}, and self combustion method \cite{r10}.  A review on the different humidity sensing mechanism and operating principle for ceramics is reported in the literature \cite{r6}.  Kamila Suri \etal reported the humidity sensing properties of  $\alpha$-Fe$_2$O$_3$ and polypyrrole nanocomposites \cite{r5}.  Tulliani \etal also reported the humidity sensing properties of  $\alpha$-Fe$_2$O$_3$ and doping effects \cite{r11}.  Most of the humidity sensors reported in literature works are at elevated temperatures.  In this paper a potential ceramic humidity sensor working at room temperature is investigated.  The structural, electrical and magnetic properties of copper, cobalt and its mixed ferrite materials (\cucox with x=$0.0$, $0.5$ and $1.0$) prepared by self combustion method are reported. These nanoceramics have been used as humidity sensors due to its porous nature created during the combustion process.
\section{Experimental Details}\label{exp}
Copper ferrite has been prepared by mixing copper nitrate, ferrous nitrate aqueous solutions with citric acid in $1:1$ stoichiometric ratio.  The pH of the solution is adjusted to $7$ using liquor ammonia.  The obtained sol was then allowed to evaporate in a beaker by keeping the solution temperature at 80-90$^o$C and it results into high viscous gel.  The resultant gel has been kept inside a preheated oven at 300$^o$C.  Within 5 to 10 minutes, a large amount of gas is evolved according to the equation \cite{r12} and the self combustion reaction has completed.
$$ C_6H_8O_7 + 6 NO_3 \rightarrow  6CO_2 + H_2O + 6OH^-  + 6NO $$
In this citric acid acts as a fuel to produce the necessary bonding with metal ions and prevents the metal to precipitate as metal hydroxides \cite{r13}.\\
X-ray diffraction pattern have been taken in X'Pert PRO diffractometer, using Cu-K$_\alpha$  radiation of wavelength 1.54 \AA\ and microstructure analysis was carried out on a scanning electron microsocope (SEM).  The Fourier transform infrared (FTIR) spectrum for the ferrite samples and gel (copper ferrite) samples are recorded using Bruker Tensor $27$ in the region of 4000 cm$^{-1}$ to 400 cm$^{-1}$.  Conductivity measurements have been carried out using two probe method from room temperature to 700$^o$C using Keithley source measure meter model 2400 on pellets having 13 mm diameter and 1 mm thickness.  The temperature is varied from 30$^o$C to 700$^o$C in steps of 25$^o$C for both heating and cooling cycles.  The magnetic properties are investigated using EG \&\ G PARC 4500(USA) vibrating sample magnetometer(VSM).  The humidity sensing behaviour of the material was measured with an indigenous set-up made of glass chamber in which relative humidity can be varied.  The compressed air dehydrated over silica gel and calcium chloride was directed into the chamber.  The humidity level is varied from 38 to 95\% by bubbling air through water and mixing it with dry air.  These ferrite samples of 13 mm diameter and 1 mm thickness are placed in between two silver electrodes in the chamber which are connected to the Keithley source meter model 2400 to measure the change in resistance with respect to relative humidity (RH).
\section{Results and discussions}\label{res}
\subsection{XRD Analysis}  
\begin{figure}\centering
\includegraphics[width=7cm]{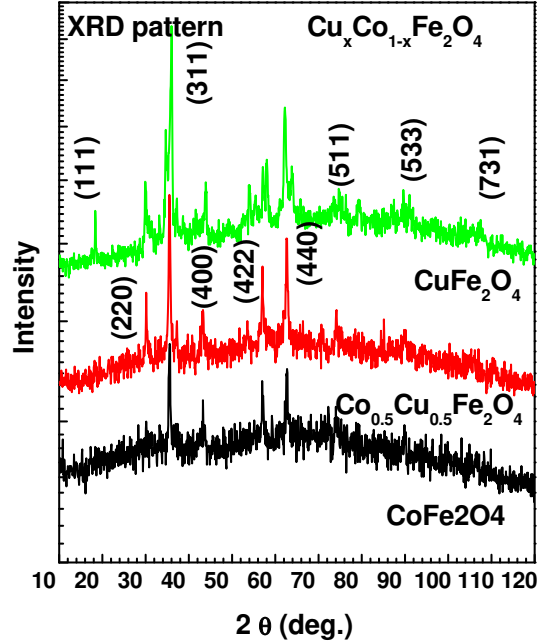}
\caption{X-ray diffraction patterns (Cu K$_\alpha$) for \cucox (x=0.0, 0.5 and 1.0) nanoparticles.}\label{fig1}
\end{figure}

\begin{figure}\centering
\includegraphics[width=7cm]{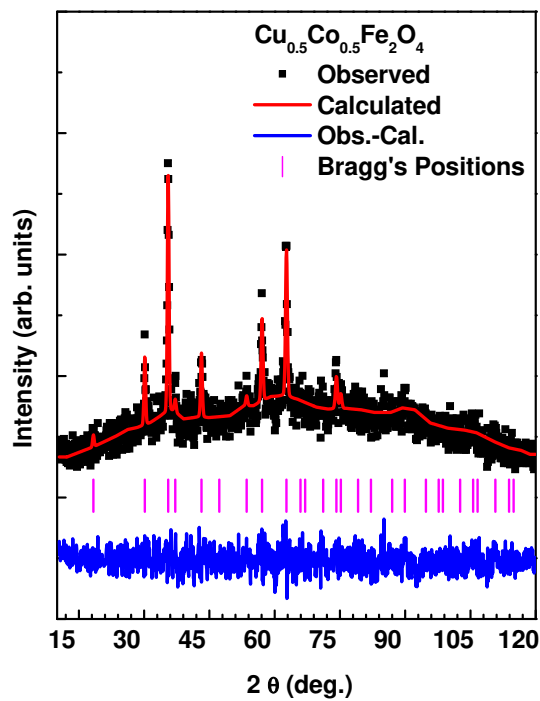}
\caption{Reitveld's fitting for XRD pattern of \cuco nanoparticles with ``goodness of fit'' $\chi^2=0.218$, Bragg's R-factor=0.442 and RF-factor=0.351. The graphs is plotted for observed points and calculated points on the upper line. Below is the difference between the two. Middle line points shows the Bragg's positions for the FD3M space group, which calculates the values of lattice constants as a=b=c=8.372 and angle=90$^o$.}\label{fig2}
\end{figure}

\Fref{fig1} shows the indexed x-ray diffraction (XRD) patterns of \cucox (x=0.0, 0.5 and 1.0). Different peaks were identified by using the JCPDS database (for copper ferrite, JCPDS No. 25 0283). XRD patterns show the formation of single phase cubic structure with dominant peak corresponding to (311) reflection indicating that the crystallites are preferentially oriented along (311) plane. The breadth of the Bragg peak is a combination of both instrument and sample dependent effects. To decouple these contributions, it is necessary to collect a diffraction pattern from the line broadening of a standard material such as silicon to determine the instrumental broadening. The instrumented corrected broadening $\beta_{hkl}$ corresponding to the diffraction peak of \cucox (x=0.0,0.5 and 1.0) was estimated by using the relation
\begin{equation}
\beta_{hkl}= \left[ \left( \beta_{hkl} \right)^2_{\rm measured} -\left( \beta_{hkl} \right)^2_{instrumental} \right]^2
\end{equation}
Using the $\beta_{hkl}$ of XRD peaks the crystallite size is calculated by Scherrer's formula
\begin{equation}
D_{hkl} = \frac{0.9\lambda}{\beta_{hkl}\cos\theta_{hkl}}
\end{equation}
Where $D_{hkl}$ = volume weighed crystallite size, $\lambda$=wavelength of CuK$\alpha$ (1.54 \AA\ ) and $\beta_{hkl}$= instrumental corrected full width at half maximun (FWHM) of peak in radian.\\
The XRD pattern of \cucox was refined by Rietveld method using FullProf suite \cite{r14}, within FD3M space group and shown in \Fref{fig2} for \cuco. The average crystallite size is calculated for the three high intense reflections using Debye-Scherrer's formula for cobalt ferrite, copper cobalt ferrite and copper ferrite are found to be 35 nm, 33 nm and 27 nm respectively. The lattice parameters are refined using PowderX \cite{r14a} calculations for cobalt ferrite, mixed copper cobalt ferrite and copper ferrite are found to be 8.381 \AA\, 8.372 \AA\  and 8.37 \AA\  respectively. The reflections (400) and (731) are excluded in refinement due to large residual values.  It is observed that with increase in Cu content the lattice constant and unit cell volume decreases.  The decrease in lattice constant and unit cell volume is due to the smaller ionic radii of the doped cation i.e. Cu$^{2+}$ (0.730 \AA\ ) than that of Co$^{2+}$ (0.745 \AA\ ). The increase in the X-ray density ($\rho_{\rm x-ray}$) is due to the increase in the molar masses of the doped sample i.e. Cu$^{2+}$ (63.55 g mol$^{-1}$) as compared to Co$^{2+}$ (58.93 g mol$^{-1}$).\\
After the addition of Cu to cobalt ferrite a shift in most intense (311)  peak is observed.  If the diffraction peak shift to the lower angles, a tensile stress can be realized, where as a shift towards higher angles indicates a compressive stress \cite{r15}. The compressive strain along [311] direction has been calculated using the following relation
\begin{equation}
\Delta d/d_{undoped} =(d_{doped} - d_{undoped})/d_{undoped}
\end{equation}
Where $\Delta d$ is the change in the d-spacing w.r.t. undoped sample (pure cobalt ferrite). The strain is  due to the substitution of Cu ion in place of Co ion as the ionic radii of Cu (0.73 \AA\ ) is less than Co (0.745 \AA\ ). The strain calculated for (311) direction is given in \Tref{tab1}.
\Table{Hysteresis loop parameters for copper ferrite, cobalt ferrite and copper-cobalt mixed ferrite.}
\begin{tabular}{llll}
\br
Material&	Ms (emu/g)&	Hc (Oe)& Mr (emu/g)\\ \mr
CuFe2O4&	20.00&	838&	11.34 \\
CoFe2O4&	6.31&	1951&	3.42\\
Cu0.5Co0.5Fe2O4&	9.06&	1047&	4.07\\ \br \end{tabular}\label{tab1}
\endtab
Since all the XRD patterns are recorded under the same experimental conditions therefore the crystalline nature of these materials can be compared by calculating  the  degree of crystallinity (Nc) by using the relation
\begin{equation}
N_c =(I_{doped} -I_{undoped})/I_{undoped}
\end{equation}
Where $I_{doped}$ is the integrated intensity when x=0.5 and 1.0, $I_{undoped}$ is the integrated intensity  when x=0 (i.e. pure cobalt ferrite). A positive value of N$_c$ indicates the improvement in the crystallinity compared with the undoped and negative value indicate the decrease in crystallinity (\Tref{tab2}).
\Table{X-ray powder diffraction data of \cucox for most intense (311) reflection.}
\begin{tabular}{lllllll} \br
Conc.(x)&	2$\theta$ (deg.)&	d$_{hkl}$ \AA\ &	FWHM(deg.)&	D$_{hkl}$(nm)&	Lattice Strain&	Crystallinity \\ \mr
x=0.0&	36.0221&	2.49332&	0.2480&	33.686&	---&	----- \\
x=0.5&	35.5226&	2.52723&	0.3306&	25.234&	0.01360&	-0.713\\
x=1.0&	35.5275&	2.52689&	0.3306&	25.234&	0.01346&	-0.475\\ \br \end{tabular} \label{tab2}
\endtab
\begin{figure}\centering
\includegraphics[width=8cm]{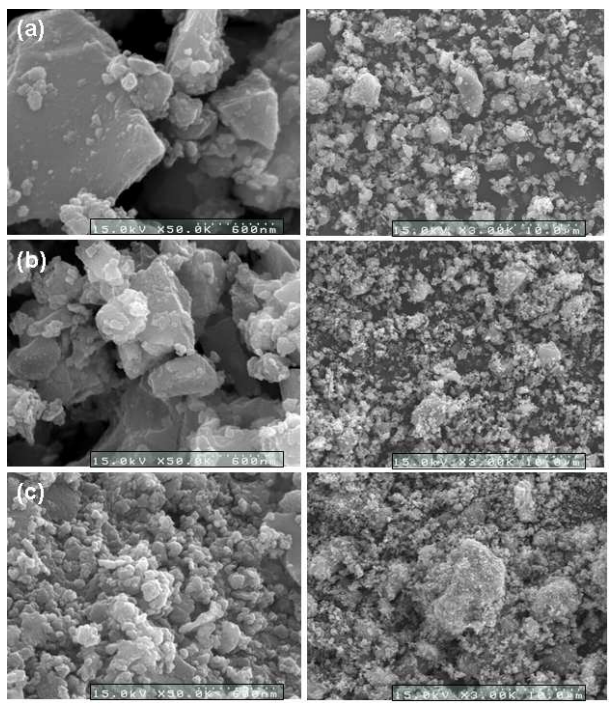}
\caption{SEM micrographs for \cucox (x=0.0, 0.5 and 1.0).(a) CoFe$_2$O$_4$ (b) \cuco (c) CuFe$_2$O$_4$, each at 600nm and 10$\mu$m scale respectively.}\label{fig3}
\end{figure}

SEM micrographs were used to see the grain micro-structure of the nanoparticles, which would provide a better view of the grain development and grain sizes. SEM micrographs are shown in \fref{fig3}(a-c) at different resolution scales. From the micrographs, it is clear that grains have also different morphologies than spherical only. The grain sizes  measured by ImageJ (1.42q) for  \cucox (x=0, 0.5 and 1.0) are 8.913, 7.095 and 10.203 $\mu$m respectively.
\subsection{FT-IR spectroscopy} 
\begin{figure}[ht!]\centering
\includegraphics[width=6cm]{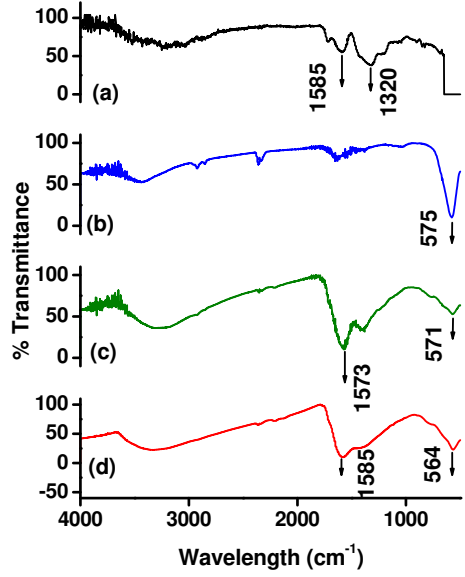}\\
\caption{(color online) FTIR spectra for (a) copper ferrite gel (b) copper ferrite (c) cobalt ferrite and (d) copper-cobalt ferrite. The band near 575 cm$^{-1}$ in (b) to (d) samples confirm the presence of ferrite phase.}\label{fig4}
\end{figure}

The FT-IR spectra of copper ferrite gel, copper ferrite, cobalt ferrite and mixed cobalt-copper ferrite samples are recorded in the range of 400-4000 cm$^{-1}$ and shown in \fref{fig4}(a-d). In the spectrum of gel, peak at 1320 cm$^{-1}$ is due to NO$_3$ vibration \cite{r16} indicating the presence of nitrate ions in the gel.  This peak is not present in the ferrite materials as seen in \fref{fig4}(b-d).  The peaks at (1573 - 1585 cm$^{-1}$) exhibit the presence of citrate ions, chemically bounded to the metal atoms \cite{r17}.  The intense bands observed at 575 cm$^{-1}$, 571 cm$^{-1}$ and 564 cm$^{-1}$ in copper ferrite, cobalt ferrite and mixed ferrite respectively. The change in band position on going from one concentration to other may be due to change in the inter-nuclear distance of Fe$^{3+}$-O$^{2-}$ in the equivalent lattice sites. These bands are attributed to the stretching vibration of  Fe$^{3+}$-O$^{2-}$  and this is the characteristic peak of ferrites \cite{r10}. This peak is not present in the copper ferrite gel and it reveals that ferrite phase is produced only after the combustion reaction.
\subsection{Conductivity studies} 
\begin{figure}[ht!]\centering
\includegraphics[width=6cm]{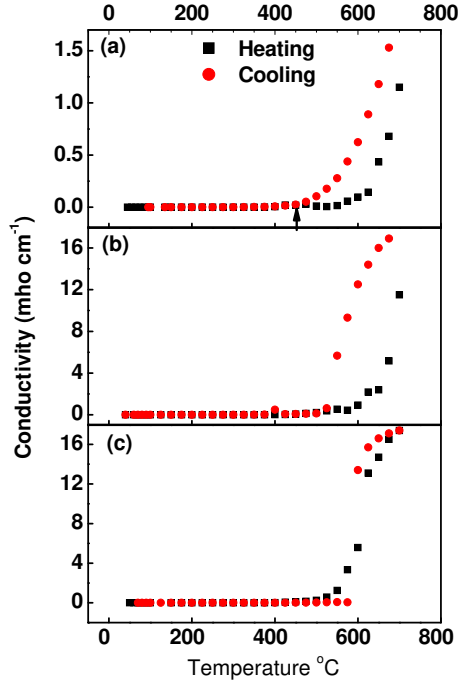}\\
\caption{(color online) Variation of conductivity with temperature for (a) copper ferrite (b) cobalt ferrite and (c) copper-cobalt ferrite.}\label{fig5}
\end{figure}

\Fref{fig5}(a-c) shows, the conductivity changes with temperature for copper ferrite, cobalt ferrite and \cuco respectively.  During the heating cycle, the conductivity of all the three samples increases as the temperature increases.  During the cooling cycle, the conductivity decreases with the fall in temperature but follows a new path that leads to a thermal hysteresis.  The observed step change in conductivity on cooling can be attributed to the defects present in the pristine material which gets smoothened out during the heating cycle.  The conduction mechanism in ferrites is explained on the basis of the Verwey de Boar mechanism \cite{r18} that involves exchange of electrons between the ions of the same element having more than one valence state.  At low temperature low conductivity is observed which may be the result of large voids and less cohesion. But the high conductivity at high temperature may be due to polaron hopping. It is reported that copper ferrite acts both as n- and p-type semiconductors \cite{r19,r20}. The two competing mechanism may be due to the hopping of electrons between Fe$^{2+}$   and   Fe$^{3+}$ ions and jumping of holes between Co$^{2+}$ and   Co$^{3+}$, and Cu$^{2+}$  and Cu$^{1+}$ as shown  in the following redox reaction:
\begin{eqnarray}
Fe^{2+}   \rightarrow Fe^{3+} + e^-, &  Co^{3+} \rightarrow Co^{2+} + e^+(hole), \nonumber\\
Fe^{2+} + Co^{3+} \rightarrow Fe^{3+} +  Co^{2+}~~~~~~~~~ ~~~~& Cu^{2+} \rightarrow Cu^{1+} + e^+, \nonumber\\
Fe^{2+} + Cu^{2+} \rightarrow Fe^{3+} +  Cu^{1+} &
\end{eqnarray}
At high temperature the fractions of Fe$^{2+}$ and availability of electrons will be much more than at low temperature.  Therefore the conduction at lower temperature is due to extrinsic type while at higher temperature is due to polaron hopping. Moreover we have observed a sharp change in the Arrhenius plot (not shown here) for all samples, which is due to the change in the conduction mechanism. The activation energy for the all sample is lower at low temperature as compared to high temperature.
In copper ferrite, \fref{fig5}(a), the heating and cooling cycle retraces the same path up to 475 $^o$C, but gets separated in the region 475 $^o$C to 700 $^o$C.  Curie temperature of copper ferrite is reported as 455 $^o$C in the literature \cite{r21}.  Chao Liu \etal, has reported that in cobalt ferrite the phase transition from ferromagnetic to paramagnetic is observed at 517 $^o$C \cite{r22} whereas in the present work the knee point is observed at 525 $^o$C.  The observed heating-cooling transition temperatures (Knee points) are 475 $^o$C, 525 $^o$C, 500 $^o$C for copper ferrite, cobalt ferrite and copper-cobalt mixed ferrite respectively.
\subsection{Magnetic Properties} 
\begin{figure}[ht!]\centering
\includegraphics[width=6cm]{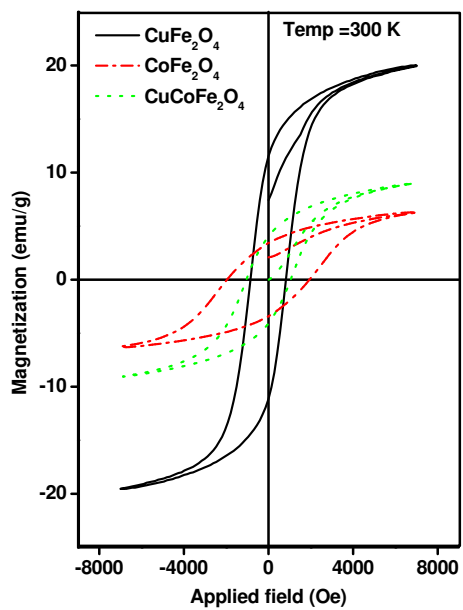}\\
\caption{(color online) B-H hysteresis loop for copper ferrite, cobalt ferrite and copper-cobalt ferrite at room temperature.}\label{fig6}
\end{figure}

\Fref{fig6} shows the hysteresis loop for copper ferrite, cobalt ferrite and mixed copper- cobalt ferrite materials recorded using vibrating sample magnetometer.  Various magnetic properties such as saturation magnetization (Ms) remanence (Mr) and corecivity (Hc)   are calculated from the hysteresis loop and given in \Tref{tab2}.  Hysterisis loop for copper ferrite and cobalt ferrite shows that these two materials have net magnetization even before applying the magnetic field.  But the copper-cobalt mixed ferrite loses this property.  Copper ferrite is known to be magnetically soft, with the low coercive values at room temperature.  Pure Cu$^{2+}$ ions are diamagnetic in nature and hence the formation of copper ferrite gives the low coercive values.  In copper ferrite, the effect of variation of crystallite size (50 - 220 nm) on saturation magnetization (39 to 47 emu/g) is reported \cite{r23}.  The observed Ms value of 20 emu/g in the present work can be linked to the decrease in crystallite size as evident from XRD analysis and it is in agreement with the work of Ahemed A. Farghali \etal \cite{r23}.  However the measured coercivity value of 838 Oe is much larger and it can be attributed to the strength of magnetic moments formed due to the self-combustion preparation technique.
The hysteresis loop for cobalt ferrite has a high coercive field of 1951 Oe with small saturation magnetization of 6.31 emu/g. Pure Co$^{2+}$ and Fe$^{3+}$ ions are ferromagnetic in nature.  So in cobalt ferrite the bonding between A (tetrahedral) and B (Octahedral) sites lead to higher coercivity multidomain structure.  Cannas \etal \cite{r10} reported that cobalt ferrite prepared by self combustion method has 65.9 emu/g magnetization and 1550.8 Oe coercive field.  Cobalt ferrite prepared by sol-gel method (800 $^o$C annealing temperature) has 2020 Oe coercivity and 76.5 emu/g Ms  as reported by Jae Gwang Lee \etal \cite{r7}.  Yu Qu \etal reported that the value varies from 3.3 emu/g to 29.5 emu/g for different annealing temperature and a maximum coercive field of 1180 Oe \cite{r24}.  The above discussion shows that the magnetization and coercive field values strongly depend on the preparation technique and temperature.
The mixed ferrite, \cuco, has intermediate values in both magnetization and coercivity. This shows that mixing of cobalt with copper ferrite increases the H$_c$ values and decreases the M$_s$ values. The addition of high coercive Co$^{2+}$ with copper ferrite, leads to the A-A (Co$^{2+}$ \& Cu$^{2+}$) interaction in the tetrahedral site and create the single domain system, which requires greater energy rather than the movement of walls (multi domain). Thus the involvement of Cu$^{2+}$ ions essentially decreases the net coercive values in mixed ferrites. The presence of Co$^{2+}$ and Cu$^{2+}$ ions at the same site has also been discussed through x-ray absorption spectroscopy elsewhere \cite{r25}.
\subsection{Humidity sensor} 
\begin{figure}[ht!]\centering
\includegraphics[width=6cm]{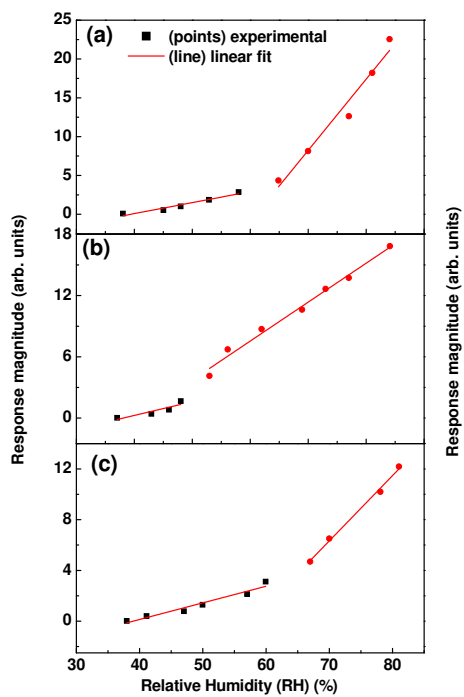}\\
\caption{(color online)Variation of response magnitude with RH for (a) copper ferrite (b) cobalt ferrite and (c) copper-cobalt ferrite.} \label{fig7}
\end{figure}

The ferrite materials are porous in nature and have surface oxygen atoms which essentially arise due to the sample preparation technique. When the material adsorbs the humidity, its resistivity decreases due to the increase of charge carriers, protons, in the ferrite and water system \cite{r26}.  The adsorption of water on the surface of the material leads to the dissociation of hydrogen ions.  These hydrogen ions bonded with the surface lattice oxygen atom, forms the hydroxyl groups \cite{r5} as shown in the equation
$$ H^+ + O_o \leftrightarrow [OH]^- $$
where O$_o$ corresponds to oxygen at lattice sites.  The hydroxyl groups thus produced are bonded with the lattice iron atoms and liberate the free electrons \cite{r27}.
$$ [OH]^- + Fe \leftrightarrow [OH - Fe] + e- $$
Thus conductivity increases with increase in humidity because of the production of free electrons.\\
\Fref{fig7}(a-c) shows the response magnitude of the copper ferrite, cobalt ferrite and copper-cobalt ferrite respectively, for various humidity ranges.  Response magnitude is defined as,
$$ {\rm Response~ magnitude} = \Delta\sigma/\sigma $$
where $\Delta\sigma$  is the change in conductivity at particular RH and $\sigma$ is the conductivity at low RH.
In copper ferrite, \fref{fig5}(a), two linear regions are noticed.  From 38 to 58 \% RH, the increase in sensitivity is slow and it is fast in the region 65 to 84 \% RH.  The total conductivity increases by 17 times in the humidity range of 38 \% to 84 \% of RH and it explains its potential use in humidity sensing.  In cobalt ferrite, \fref{fig7}(b), the sensitivity varies linearly as two regions from 37 to 48 \% RH and 53 to 84 \% RH.  Copper-cobalt mixed ferrite, \fref{fig5}(c) has also two linear regions, one in the range of 38 to 60 \% RH and the other in 67 to 81 \% RH.  In comparison, at 80 \% RH, the copper ferrite, cobalt ferrite and copper-cobalt mixed ferrite materials show the response of 17.05, 15.12 and 11.70 respectively.
\section{Conclusions}\label{conclusions}
Copper ferrite, cobalt ferrite and copper-cobalt mixed ferrite nanomaterials were prepared by self combustion method. X-ray diffraction pattern shows the crystalline nature of the materials and these nano-crystallites are preferentially oriented along (311) plane.  FTIR spectrum of bulk material shows the characteristic peak of ferrites. The temperature variation of the electrical conductivity of all the samples shows a thermal hysteresis and definite break in conductivity, which corresponds to ferrimagnetic-paramagnetic transition. The activation energy in the paramagnetic region is higher than in the ferrimagnetic region. VSM studies revealed that substitution of cobalt with copper ferrite increases the corecivity and decreases the saturation magnetization.  Humidity sensing properties are studied and response magnitude indicates that copper ferrite has maximum sensitivity.
\ack
The authors acknowledge the Inter-University Accelerator Centre, New Delhi for financial support (IUAC - UFUP 41334) and Korea Institute of Science and Technology (KIST- 2V01450).

\end{document}